\documentclass[pra,showpacs,twocolumn,showkeys]{revtex4-1}
\usepackage{graphics}
\usepackage{amsmath}
\usepackage{color}
\usepackage{enumerate}

\begin{document}

\title{The paradox of the many-state catastrophe of fundamental limits and the three-state conjecture}
\author{Shoresh Shafei and Mark G. Kuzyk}
\affiliation{Department of Physics and Astronomy, Washington State University, Pullman, Washington  99164-2814}

\begin{abstract}
The calculation of the fundamental limits of nonlinear susceptibilities posits that when a quantum system has a nonlinear response at the fundamental limit, only three energy eigenstates contribute to the first and second hyperpolarizability.  This is called the three-level ansatz and is the only unproven assumption in the theory of fundamental limits.  In light of the observation that the measured nonlinear response of a majority of molecules falls far short of these limits, the three-level ansatz warrants closer scrutiny.  All calculations that are based on direct solution of the Schr\"{o}dinger equation, including systems of arbitrarily-placed electrons and nuclei in external electromagnetic fields and interacting electrons in an arbitrary potential, whose configuration space is sampled using numerical optimization techniques, yield intrinsic hyperpolarizabilities less than 0.709 and intrinsic second hyperpolarizabilities less than 0.6.  In this work, we show that relaxing the three-level ansatz and allowing an arbitrary number of states to contribute leads to divergence of the optimized intrinsic hyperpolarizability in the limit of an infinite number of states - what we call the many-state catastrophe.  This is not surprising given that the divergent systems are most likely not derivable from the Schr\"{o}dinger equation, yet obey the sum rules.  The sums rules are the second ingredient in limit theory, and apply also to systems with more general Hamiltonians.  These ``exotic Hamiltonians" may not model any real systems found in nature.  Indeed, a class of transition moments and energies that come form the sum rules do not have a corresponding Hamiltonian that is expressible in differential form.  In this work, we show that the three-level ansatz acts as a constraint that excludes many of the nonphysical Hamiltonians and prevents the intrinsic hyperpolarizability from diverging.  We argue that this implies that the true fundamental limit is smaller than previously calculated.  Since the three-level ansatz does not lead to the largest possible nonlinear response -- contrary to its assertion -- we propose the intriguing possibility that the three-level ansatz is true for any system that obeys the Schr\"{o}dinger equation, yet this assertion may be unprovable.

\end{abstract}
\pacs{42.65.An, 33.15.Kr}

\maketitle

\section{Introduction}
The fundamental limits of the first and second hyperpolarizabilities\cite{kuzyk00.01,kuzyk00.02,kuzyk03.02,kuzyk03.01} can be used to define scale-invariant quantities\cite{kuzyk10.01,kuzyk13.01} that have been used to test the theoretical basis of the nonlinear susceptibility as well as exploring their consequences as embodied in real quantum systems. Slepkov \emph{et al.} used the theory of fundamental limits to understand scaling of the surprising NLO properties of polyene oligomers \cite{slepk04.01}.  May \emph{et al.} used the fundamental limits as an absolute standard for comparing molecules, independent of their size, to show the promise of small molecules.\cite{May05.01,May07.01} Chen \emph{et al.} showed that the nonlinear response of certain nano-engineered polymers gave a larger intrinsic nonlinear-optical response than what would be expected from the individual building blocks\cite{wang04.01}. The first- ($\beta$) and second-order ($\gamma$) response of twisted $\pi$-electron chromophores were identified to constitute a new paradigm for enhanced electro-optic materials\cite{Kang05.01,Kang07.01,He11.01} based on the fact that they fell far into the gap between the best molecules every measured and the fundamental limit.\cite{zhou08.01} The fundamental limits have also been used to define the concept of modulated conjugation in the bridge between the donor and acceptor ends as a new paradigm for enhancing the \emph{intrinsic hyperpolarizability}, $\beta_{int}$\cite{perez07.01,perez09.01} defined by
\begin{equation}\label{betaIntrinsic}
\beta_{int} = \frac{\beta}{\beta_{max}},
\end{equation}
where $\beta_{max} $ is the fundamental limit\cite{kuzyk00.01,kuzyk00.02,kuzyk03.02,kuzyk03.01} of the first hyperpolarizability and is calculated using the sum rules\cite{thom25.01,reich25.01,kuhn25.01} and the three-level ansatz.\cite{kuzyk00.01,kuzyk00.02,kuzyk03.02,kuzyk03.01}

While the fundamental limits have been used as a metric in comparing molecules, provide insights in understanding the scaling behavior of molecular homologues, and act as a guide in developing new molecular paradigms, the assumptions underlying the limits have not been scrutinized, nor have their implications been fully studied.  This paper revisits the assumptions and investigates their consequences to fundamental physics.  The paper is organized into four parts.  First, the derivations of the limits are reviewed and the assumptions fully interpreted.  Next, the behavior predicted from the limits is compared with theoretical calculations and experimental measurements of the nonlinear susceptibility.  The third section demonstrates how the theory of the fundamental limits changes when the assumptions are relaxed, and their implications discussed.  Finally, the last section describes hints of potentially new fundamental physics that lies just beyond our horizon of understanding.

\section{Fundamental limits}\label{sec:limits}

The fundamental limits are calculated in three straightforward steps.  First, the sum-over-states expressions of Orr and Ward\cite{orr71.01} for the nonlinear susceptibilities are simplified using the generized Thomas-Reiche-Kuhn sum rules -- which relate the position operator matrix elements, $x_{ij} \equiv \left<i\left|x\right|j\right> $, and energies, $E_i$, to each other,
\begin{equation}\label{sumRules}
\sum_{n}\left(E_n - \frac{E_m + E_p}{2}\right)x_{mn}x_{np} = \frac{\hbar^2 N}{2m}\delta_{mn} ,
\end{equation}
where $n$, $m$ and $p$ index the energy eigenstates, and $N$ the number of electrons.  Then, the three-level ansatz is applied under the assumption that a three-level model accurately describes any system with a nonlinear-optical response near the fundamental limit.  This leads to an expression with only two adjustable parameters, the ratio between the energies of the two states, $E = E_{10}/E_{20}$, and the matrix element of the position operator, $x = x_{10}/x_{10}^{max}$, where $x_{10}^{max}$ is the upper bound of $x_{10}$, and it is determined from sum rules.  Finally, the expression for the nonlinear susceptibility is optimized under the assumption that $x$ and $E$ are independent.  The process can be reversed, with the 3-level ansatz being applied prior to the sum rules, as we outline below.

First, we reiterate the definition of the Three-Level Ansatz (TLA):
\begin{quote}
{\emph{when the hyperpolarizability of a quantum system is at its fundamental limit, only three states contribute to the nonlinear response.}}
\end{quote}
The converse is not true; that is, a quantum system in which only three three states contribute to the nonlinear response will not necessarily have a a large nonlinear response and may in fact be far from the limit.

The three-level ansatz is motivated by the simpler calculation of the off-resonance polarizability, given by,
\begin{eqnarray}\label{polarizability}
\alpha && = 2 \sum_{n} \frac { e^2 \left|x_{n0}\right|^2} {E_{n0}} = 2 \sum_{n} \frac { e^2  \left|x_{n0}\right|^2 E_{n0}}  {E_{n0}^2} \nonumber \\
&& \leq \frac{2e^2 } {E_{10}^2} \sum_{n} \left|x_{n0}\right|^2 E_{n0} .
\end{eqnarray}
Using Eq. (\ref{sumRules}) with m=p=0,
\begin{equation}\label{polarizabilityMax}
\alpha \leq \frac{e \hbar^2} {m} \frac{ N}{E_{10}^2} \equiv \alpha_{max}.
\end{equation}
An alternative method for getting Eq. (\ref{polarizabilityMax}) is to note that Eq. (\ref{polarizability}) is a sum of positive definite terms.  If the full transition strength is placed in State 1, the sum rules predict that all other transition moments $x_{n0}$ vanish.  Thus, since the numerator is made maximum by placing all the transition strength between states 0 and 1, and this denominator is the smallest of all terms, then $\alpha$ is maximum in a two-level model.

Consequently, one might expect that the same may be true for $\beta$; but, two problems arise.  First, since the sum over states expression does not have all positive-definite terms, it is not a simple task to show that placing all the transition strength in the first excited state is the best strategy.  A more damning problem is that the sum rule for the two-level model with $m=0$ and $p=1$ yields
\begin{equation}\label{two-level-contra}
E_{10} x_{10} \left( x_{11} - x_{00} \right) = 0,
\end{equation}
which does not allow both $x_{10}$ and $ x_{11} - x_{00}$ to be nonzero.  Since in the two-level model, $\beta \propto \left| x_{10} \right|^2 \left( x_{11} - x_{00} \right)$, the implication is that $\beta=0$ for a two-level system -- clearly not a maximum.  However, no such problems arise for the three-level model.  Since using the minimum possible number of states makes the transition strengths of the non-vanishing terms large, and if the transition energies are small -- that is, the lowest energy states are the dominant ones, then the three-level model will yield the largest nonlinear response.  Optimization of this expression by varying the independent physical quantities is then assumed to lead to the fundamental limit.

Since this sum rule derivation of limits appears to hold for all measurements and all exact analytical calculations, the three-level approximation for an optimized system appears reasonable.  Indeed, the highly successful two-level model, used in understating the nonlinear response of large numbers of donor/acceptor molecules by hundreds of researchers,\cite{kuzyk98.01} has not been questioned because of its success despite the fact that such an expression violates the sum rules.  The two-level model may work well for molecules with a hyperpolarizability far from the fundamental limit, where many states contribute but two dominate the response, perhaps due to slight resonance enhancement.  In contrast, when the two-level model is constrained to obey the sum rules and forcing only two states to contribute while suppressing all others, the two-level model becomes unphysical.

Applying the sum rules to a three-level model yields its own set of problems.  One can show that the truncation of the sum rules is perfectly legitimate when applied to a few-state sum-over-states (SOS) expression of the hyperpolarizability in the manner used in calculating the limits.\cite{kuzyk06.03}  The problem arises in the choice of sum rules used in simplifying the SOS expression.  For example, in a three-level model, the largest state index is 2 and the sum rule corresponding to Eq. (\ref{sumRules}) with $m=p=2$ (call this sum rule $\Sigma_{22}$ ) is nonsensical and yields the obviously nonsensical result that,
\begin{equation}\label{sumRules22}
- \sum_{n} E_{2n} \left|x_{2n} \right|^2 = \frac{\hbar^2 N}{2m},
\end{equation}
that is, a negative number equals a positive one.

In calculating the limits, the sum rule equation $\Sigma_{22}$ is ignored.  Using the sum rules $\Sigma_{00}$, $\Sigma_{10}$ and $\Sigma_{20}$ reduces the three-level approximation of the SOS expression to one with two adjustable parameter that are varied to find the maximum.  The assumption here is that all nonlinear-optical susceptibilities near the limit can be approximated using this two-parameter model.

The above approach ignores sum rule $\Sigma_{12}$.  One may argue that this choice is arbitrary, as did Champagne and Kirtman.\cite{Champ05.01} For example, why not also ignore sum rule $\Sigma_{20}$.  With this choice, there would be no upper bound on the hyperpolarizability.  In the end, the appropriate choice is the one that is consistent with the data.  Potential optimization studies,\cite{zhou06.01,zhou07.02,szafr10.01} nuclear placement,\cite{kuzyk06.02,watkins09.01} application of electromagnetic fields,\cite{watkins09.01} and interactions between electrons in a potential well\cite{watkins11.01} all show maximum hyperpolarizabilities of $\beta_{int} \leq 0.709$ and $\gamma_{int} \leq 0.6$.\cite{watkin12.01}  Using fewer sum rule equations makes the upper bound infinite while additional constraints decreases the limit below these values.  Thus, while not a rigorous proof, these observations show that the approximations used in calculating the limit are reasonable.

In summary, the calculations of the fundamental limits uses the three-level anzatz, which has not been rigourously derived, but appears to hold over a broad set of observations.  Similarly, using the sum rules $\Sigma_{00}$, $\Sigma_{10}$ and $\Sigma_{20}$, and ignoring the others is somewhat arbitrary but again yields the correct results.  These assumptions are scrutinized and their implications discussed below.

\section{Comparison of Limits with Data}

The TLA and the neglect of the pathological sum rules are the two untested assumptions of the theory of limits.  Since they cannot be tested directly, and proof with analytical techniques has been unsuccessful, an indirect approach is to sample a large enough domain of the full configuration space of quantum systems to  determine if any of the predictions of the theory are violated.  Are there instances where a system beats the limits?  Do all system fall substantially short of the limit? How many states are typically involved near the limits?  These and related questions can be answered using analytical calculations and experimental data.

Since the hyperpolarizability is not a scale-invariant quantity, we use the \emph{intrinsic hyperpolarizability}, $\beta_{int}$, as an absolute standard for evaluating a quantum system that is independent of the quantum unit's size,\cite{zhou06.01,zhou08.01,kuzyk10.01} and spans the range $[-1,1]$ with $\left|\beta_{int}\right| =  1$ defining the fundamental limits.  The factor of 30 gap between the largest nonlinearities observed (prior to 2007) by experiment  and the fundamental limits and the fact that the hyperpolarizability of the clipped harmonic oscillator is less than a factor of 2 smaller than the limit suggested that new paradigms for making better molecules and artificial quantum systems would be required to make larger intrinsic hyperpolarizabilities practical.\cite{tripa04.01}  More importantly, these studies established that the fundamental limits held for all known molecules and therefore represented the true upper bound.

To understand the nature of this gap, Tripathy \emph{et al.} used linear spectroscopy, Raman spectroscopy, $\beta$ values measured by Hyper-Rayleigh scattering and Stark spectroscopy to determine which parameters are critical.\cite{tripa04.01}  Included were tests of dilution effects due to vibronic states, investigations of unfavorable energy spacing of the molecule or atom, simplifications inherent in local field models, and an analysis of the effects of truncation of the sum rules. These studies concluded that the energy spectrum of real systems compared with the ideal is the most likely factor that keeps the hyperpolarizabilities of real molecules well below the limit.\cite{tripa04.01}

To gain insights about the gap, the effects of several different parameters -- including molecular geometry, external electromagnetic fields and electron-electron interactions -- have been investigated. The effect of molecular geometry on the hyperpolarizability is determined by varying the positions and magnitudes of charges in 2-D and correlating them with dipolar charge asymmetry and the variations of angle between point charges in octupolar structures. It was shown that the best dipolar and octupole-like molecules have intrinsic hyperpolarizabilities near 0.7.\cite{kuzyk06.02}

Hamiltonians of the standard form, given by
\begin{equation}\label{Hamiltonian}
H = \frac{p^2}{2m} + V(r),
\end{equation}
are used to calculate the sum rules. However, more general Hamiltonians also obey the sum rules.  For example, the first hyperpolarizability of molecules was studied in the presence of an external electromagnetic field modeled by a vector potential term added to the momentum, and also led to best intrinsic hyperpolarizabilities of about 0.7 even when the external field is comparable with internal molecular fields.\cite{watkins09.01} Thus, all real 1-electron systems modeled appear to have an upper bound of 0.709.

More recent work addresses the role of electron interactions on $\beta_{int}$\cite{watkins11.01} and show that while $\beta$ for two non-interacting electrons are twice that of a single-electron system, since the fundamental limit for a two-electron system scales as $N^{3/2}$,\cite{kuzyk00.01} increasing the number of electrons that do not interact with each other reduces $\beta_{int}$.  At best, the theory of fundamental limits predicts that interacting electrons will lead to $N^{3/2}$ scaling of the hyperpolarizability.  Calculations of two interacting electrons confirm this prediction, so the best values of $\beta_{int}$ for this system are again found to be about $0.7$.\cite{watkins11.01}  There is no reason to believe that adding additional electrons will lead to a larger {\it intrinsic} nonlinear response.

Other studies aimed at investigating the gap have used numerical optimization techniques to find the potential energy functions that maximize the hyperpolarizability. In this approach, many classes of starting potentials (polynomial, trigonometric, etc.) are varied until the hyperpolarizabilities are optimized using a finite-difference method to calculate $\beta$ using,
\begin{equation}\label{s-finiteDifferentMethod}
\beta = \frac{1}{2} \frac{\partial^2 p}{\partial E^2}\left|\right._{E=0},
\end{equation}
where $p$ is dipole moment calculated in the presence of the applied electric field. These too find the largest $\beta_{int}$ values to be approximately 0.71.\cite{zhou06.01,zhou07.02}

Atherton \emph{et al.} used a different approach to optimize a piecewise linear potential function of the form $V(x) = Ax_n + B_n$ for $n \in \{1,2,3,\cdots, N-1\}$, with $N$ as the number of segments.\cite{ather12.01} They also find the universal value of optimized hyperpolarizability to be $\beta_{int} \approx 0.71$.\cite{zhou06.01,zhou07.01,zhou07.02}. The authors use a Hessian matrix \cite{pres07.01} of $\beta_{int}$ to find the relevant parameters defining the hyperpolarizability and find that (a) two parameters suffice to find the appropriate potential function and b) increasing the number of parameters does not improve upon the value of $\beta_{int}$. A summary of the largest attainable $\beta_{int}$ vales are presented in Table \ref{tab:betaInt's}.

In the above approaches, the Shr\"{o}dinger equation is solved directly for many Hamiltonians. All quantum systems  with an intrinsic hyperpolarizability near the fundamental limit, independent of the underlying Hamiltonian, are found to share universal properties: the three-level ansatz is obeyed; the ratio of the second to first excited state energies is $E = E_{10}/E_{20} \simeq 0.49$; and, $X = x_{01}/x_{01}^{max} \simeq 0.79$, where $x_{01}^{max} = \hbar/(2mE_{10})^{1/2}$ is the largest possible transition moment to the first excited state. It is intriguing that such a broad range of systems share the same properties, and the natural question is why.  In addition to the fundamental questions that arise, universal properties can be used as a guide for optimizing the hyperpolarizability.

A larger configuration space can be probed using the sum rules directly in lieu of the Shr\"{o}dinger equation.   In this approach, Monte Carlo simulations are used to generate a distribution of $\beta_{int}$ values based on millions of random samplings of the energies and transition moments under the constraint that the eigenenergies, $E_i$, and transition moments, $x_{ij}$, obey the sum rules. The largest hyperpolarizabilities are found once again to be characterized by three dominant states; but, the hyperpolarizabilities can approach arbitrarily closely to the limit.\cite{kuzyk08.01,shafe10.01} Since the sum rules cover a broader domain than the Shr\"{o}dinger equation, it is not surprising that values greater than given by the  Shr\"{o}dinger equation are found.  However, none are larger than $\beta_{int} = 1$.

\begin{table}
\caption{The largest intrinsic hyperpolarizabilities obtained using various theoretical approaches}
\begin{tabular}{| l || c |}
  \hline
  \textbf{Theoretical Models} & $\mathbf{\beta_{int}}$ \\
  \hline
  Charge Asymmetry  & $ < 0.7$ \\ \hline
  External Electromagnetic Field & $\simeq 0.7$ \\ \hline
  Electron Interactions & $0.709$ \\ \hline
  Potential Optimization & $0.709$\\ \hline
  Monte Carlo Simulation & $\simeq 1$ \\
  \hline
\end{tabular}\label{tab:betaInt's}
\end{table}

The present work seeks to investigate the domain not included in the Schr\"odinger equation but constrained by the sum rules. An ancillary goal is to revisit the validity of the calculations of the fundamental limits and their applicability to systems described by standard Hamiltonians such as those given by Eq. (\ref{Hamiltonian}). Since the sum rules encompass a broader range of systems than those derivable from typical Hamiltonians,\cite{shafe10.01} we also investigate nonstandard Hamiltonians that may lead to artificial materials with larger nonlinear response. Finally, we discus the many-state catastrophe, and why the three-level ansatz may be appropriate.  This approach allows us to contrast our new findings with previous studies that lead to new interpretations of the TLA.

\section{The hyperpolarizability of N-level model}

In this section, we introduce the sum rules and use them as a constraint to optimize $\beta$ for a general 4-level model.  Subsequently we generalize those results to an N-level model and show that a highly-degenerate energy spectrum appears to break the fundamental limit.  In the infinite-state model, the limits are found to diverge.  This leads to a reformulation of the fundamental limits based on the Schr\"odinger equation, which suggests a reduced limit.

\subsection{Sum rules}

Sum rules, given by Eq. (\ref{sumRules}), are the foundation of the theory of fundamental limits of the first and second order hyperpolarizabilities, $\beta$\cite{kuzyk00.01} and $\gamma$\cite{kuzyk00.02} respectively.  They have been used to calculate the fundamental limits of the off-diagonal components of $\beta$ as measured with hyper-Rayleigh scattering,\cite{kuzyk01.01} limits of two photon absorption cross sections,\cite{kuzyk03.03} and used to formulate the dipole-free sum-over-states expression for $\beta$\cite{kuzyk05.02} and $\gamma$\cite{perez01.08}.

\subsection{4-Level Model}

In this section we will first apply sum rules to a 4-level model of the hyperpolarizability to a very specific state, namely to  what we call a fully degenerate state, and find that the hyperpolarizability can exceed the fundamental limit so the intrinsic hyperpolarizability exceeds unity.  This calculation is then generalized to an $N$-level fully degenerate state using the same method, which shows a divergence of the fundamental limit as $N \rightarrow \infty$, which we call the many-state catastrophe.

Using the Dipole-Free SOS expression,\cite{kuzyk05.02} which is a simplification derived using a subset of the sum rules, the electronic first hyperpolarizability for a 4-level model in the off-resonant regime, $\beta \equiv \beta_{4L}^{DF}$, is given by
\begin{eqnarray}\label{beta4L}
\beta_{4L}^{DF} &=& {\sum_{m \neq n}^{3}}' \beta_{mn}^{DF} \nonumber \\
&=& \beta_{12} + \beta_{21} + \beta_{13} + \beta_{31} + \beta_{23}+ \beta_{32}\nonumber \\
&=& -3e^3 \left[ x_{01}x_{12}x_{20}\left( \frac{2}{E_{10}E_{20}} - \frac{2E_{10} - E_{20}}{E_{20}^3}- \right. \right. \nonumber \\
&& \left. \left. - \frac{2E_{20} - E_{10}}{E_{10}^3} \right) + x_{01}x_{13}x_{30} \left( \frac{2}{E_{10}E_{30}} -  \right. \right. \nonumber \\
&& \left. \left.  \frac{2E_{10} - E_{30}}{E_{30}^3} - \frac{2E_{30} - E_{10}}{E_{10}^3} \right) + x_{02}x_{23}x_{30} \right. \nonumber \\
&& \left. \left( \frac{2}{E_{20}E_{30}} - \frac{2E_{20} - E_{30}}{E_{30}^3} - \frac{2E_{30} - E_{20}}{E_{20}^3} \right)\right]
\end{eqnarray}
where the prime on the sum indicates that the ground state (labeled by zero) is excluded from the summation and,
\begin{eqnarray}\label{betamn}
\beta_{mn}^{DF} &=& -3e^3 x_{0m}x_{mn}x_{n0}\left(\frac{1}{E_{n0}E_{m0}} - \frac{2 E_{n0} - E_{m0}}{E_{m0}^3}\right). \nonumber \\
\end{eqnarray}

Since the sum rules relate transition moments and energies to each other, they can be used to reduce the number of parameters in a truncated SOS expression.  It has been shown that this procedure can be applied in a way that avoids pathologies.\cite{kuzyk06.03}

We begin by using the sum rules to eliminate $x_{03}$, $x_{13}$ and $x_{23}$, as follows. The sum rule $(m,p)=(0,0)$ (see Eq. (\ref{sumRules})) for a 4-level model yields,
\begin{equation}\label{00SR}
E_{10}|x_{01}|^2 + E_{20}|x_{02}|^2 + E_{30}|x_{03}|^2 = \frac{\hbar^2}{2 m};
\end{equation}
$(m,p)=(1,1)$ gives
\begin{equation}\label{11SR}
E_{01}|x_{01}|^2 + E_{21}|x_{12}|^2 + E_{31}|x_{13}|^2 = \frac{\hbar^2}{2 m};
\end{equation}
and $(m,p)=(2,2)$ yields,
\begin{equation}\label{22SR}
E_{02}\left|x_{02}\right|^2 + E_{12}|x_{12}|^2 + E_{32}|x_{23}|^2 = \frac{\hbar^2}{2 m}.
\end{equation}
Solving Eqs. (\ref{00SR}-\ref{22SR}) for $x_{03}$, $x_{13}$ and $x_{23}$ yields,
\begin{eqnarray}
x_{03} &=& \pm \sqrt{F} x_{01}^{max} \left(1 - X^2 - \frac{Y^2}{E}\right)^{1/2}\label{x03} \\
x_{13} &=& \pm \sqrt{\frac{F}{1-F}} x_{01}^{max} \left(1 + X^2 - \frac{1-E}{E} Z^2 \right)^{1/2}\label{x13} \\
x_{23} &=& \pm \sqrt{\frac{F}{E-F}} x_{01}^{max} \left(E + Y^2 + \left(1-E\right)Z^2\right)^{1/2}\label{x23}
\end{eqnarray}
where
\begin{eqnarray}\label{EFXYZ}
E & \equiv & \frac{E_{10}}{E_{20}}, \quad F \equiv \frac{E_{10}}{E_{30}}, \quad X \equiv \frac{x_{10}}{x_{01}^{max}},\nonumber \\
&&\quad Y \equiv \frac{x_{20}}{x_{01}^{max}} , \mbox{\vspace{1em} and} \quad Z \equiv \frac{x_{12}}{x_{01}^{max}}.
\end{eqnarray}
$x_{01}^{max}$ is the largest possible transition moment from the ground state to any other state and it is given by
\begin{equation}\label{xMax}
x_{01}^{max} = \left(\frac{\hbar^2 N}{2 m E_{10}}\right)^{1/2}.
\end{equation}

Rewriting Eq. (\ref{beta4L}) in terms of $E, F, X, Y$ and $Z$ gives
\begin{eqnarray}\label{beta4L-2}
\beta_{4L}^{DF} &=& \beta_{3L}^{max}\, \mathcal{L}\left(X,Y,Z,E,F\right) ,
\end{eqnarray}
where we have used the fundamental limit of the first hyperpolarizability obtained from the three-level model,
\begin{equation}\label{beta3LMax}
\beta_{3L}^{max} = 3^{1/4} \left(\frac{e^3 \hbar^3}{m^{3/2}}\right)\left[\frac{ N^{3/2}}{E_{10}^{7/2}}\right].
\end{equation}
$\mathcal{L}\left(X,Y,Z,E,F\right)$ is given by
\begin{widetext}
\begin{eqnarray}\label{L}
&&\mathcal{L}\left(X,Y,Z,E,F\right) = \left(\frac{3}{4}\right)^{3/4} \left( X Y Z \left( 2 E+ \left(1-2E\right)E^2 -2/E + 1\right) + F\left(1 - X^2 - \frac{Y^2}{E}\right)^{1/2} \left[\frac{X}{\left(1-F\right)^{1/2}}  \right. \right. \nonumber \\
&& \left. \left. \left(1 + X^2 - \frac{1-E}{E} Z^2 \right)^{1/2}\left(2F + \left(1-2F\right)F^2 +  -2/F + 1\right) + \frac{Y}{\left(E-F\right)^{1/2}} \left(E + Y^2 + \left(1-E\right)Z^2\right)^{1/2} \right. \right.\nonumber \\
&& \left. \left. \left( \left(E+F\right)^2 - \left(\frac{2F^3}{E} + \frac{2E^3}{F}\right) \right)\right] \right) . \nonumber\\
\end{eqnarray}
\end{widetext}
$\mathcal{L}\left(X,Y,Z,E,F\right)$ in fact is the intrinsic hyperpolarizability for a 4-level model, i.e.
\begin{equation}\label{beta_intVsL}
\beta_{int}^{4L} \equiv \frac{\beta}{\beta_{3L}^{max}} = \mathcal{L}\left(X,Y,Z,E,F\right).
\end{equation}
The maximum of Eq. (\ref{beta_intVsL}) corresponds to the fundamental limit calculated from a 4-level ansatz.

Based on the theory of the fundamental limits and the three-level ansatz, all intrinsic hyperpolarizabilities should be in the range $[-1,1]$. By numerically finding the maximum of the function given by Eq. (\ref{beta_intVsL}), we find that the largest value of $\beta_{int}$ results from the extreme case where States 1 and 2 are degenerate and the energy of State 3 approaches infinity (Fig. \ref{fig:4,N}-a), or
\begin{equation}\label{ExtremeCondition}
E \simeq 1, \qquad  \mbox{and} \qquad F \rightarrow 0
\end{equation}
yielding,
\begin{equation}\label{betaInt4LMax}
\left|\beta_{int}^{4L}\right|^{max} \simeq 1.28,
\end{equation}
which implies that
\begin{equation}\label{beta4LMax}
\beta_{4L}^{max} = 1.28\, \beta_{3L}^{max}.
\end{equation}

\begin{figure}
  \includegraphics{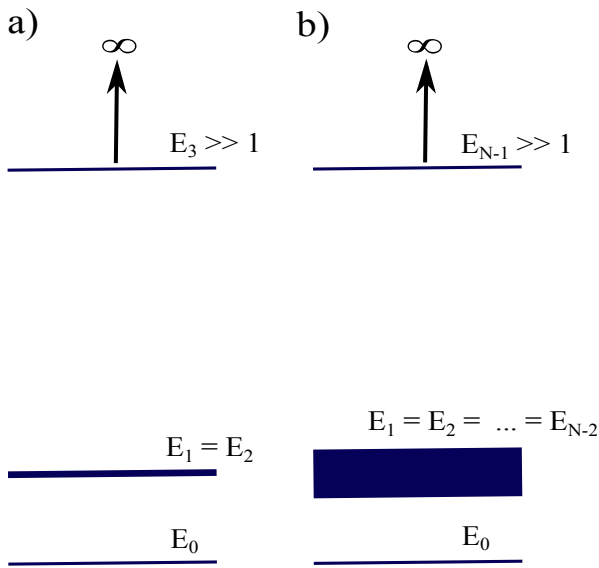}\\
  \caption{The ideal energy spectrum configurations that maximizes the hyperpolarizability of a) 4- and b) N-level models}
  \label{fig:4,N}
\end{figure}

\begin{figure}
  \includegraphics{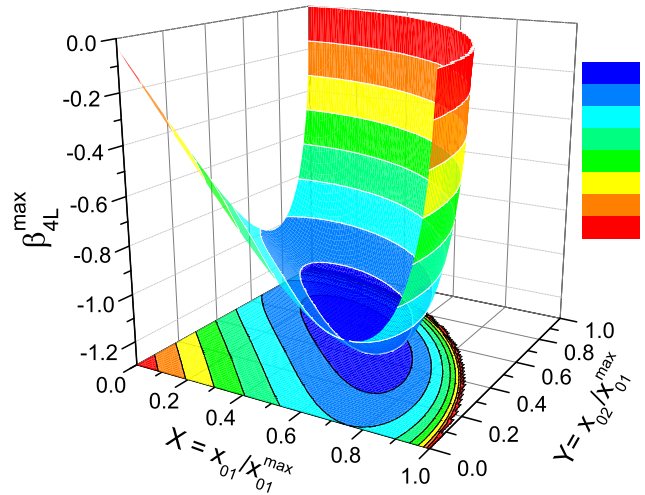}\\
  \caption{The numerical simulation for $\beta_{int}$ of a four-level model. $\beta_{int}$ exceeds the three-level maximum hyperpolarizability in the blue region.}
  \label{fig:4L}
\end{figure}
Later we will generalize the result and discuss its pathologies.
\section{N-Level Model}
In this section, we derive a general relationship for $\beta_{int}$ for an arbitrary number of eigenstates that form a fully degenerate state. Starting from the dipole free SOS expression,\cite{kuzyk05.02}
\begin{equation}\label{DF}
\beta_{int} = \left(\frac{3}{4}\right)^{3/4} {\sum_{i \neq j}}' \xi_{i0}\xi_{ij}\xi_{j0}\left( \frac{1}{e_i e_j} - \frac{2e_j-e_i}{{e_i^3}}\right) ,
\end{equation}
where normalized transition moments and energies are defined by,
\begin{equation}\label{xi}
\xi_{ij} = \frac{x_{ij}}{x_{01}^{max}}
\end{equation}
and
\begin{equation}\label{e}
e_i = \frac{E_{i0}}{E_{10}} .
\end{equation}
Monte Carlo calculations and empirical trial and error numerical simulations are used to find the maximum.   Fig. \ref{fig:4,N}-b shows the optimum, where states 1 to $(N-2)$ are degenerate,
\begin{equation}
e_1 = e_2 = e_3 = \cdots = e_{N-2},
\end{equation}
all the transition moments from the ground state to the $(N-2)$ excited states are the same, given by
\begin{equation}
\xi_{01} \simeq \xi_{02} \simeq \xi_{03} \simeq \cdots \simeq \xi_{0,N-2},
\end{equation}
and the energy of the highest-lying excited state approaches infinity,
\begin{equation}
e_{N-1} >> 1.
\end{equation}

The fully degenerate N-level state, defined by the above conditions as showed in Fig. \ref{fig:4,N}-b, yields an intrinsic hyperpolarizability of
\begin{eqnarray}\label{betamax}
\beta_{int}^{max} &=& \beta_{1,N-1} + \beta_{2,N-2} + \cdots + \beta_{N-2,N-1} + \beta_{N-1,N-2} \nonumber \\
&=& \left(\frac{3}{4}\right)^{3/4} {\sum_{i}}' \xi_{0i}\xi_{i, N-1}\xi_{N-1,0} \nonumber \\
&& \left( \frac{2}{e_i e_{N-1}} - \frac{2e_{N-1}-e_i}{{e_i^3}} - \frac{2e_i-e_{N-1}}{{e_{N-1}^3}}\right)\nonumber \\
&=& \left(\frac{3}{4}\right)^{3/4}(N-2) \xi_{10}\xi_{1,{N-1}}\xi_{N-1,0}\nonumber \\
&& \left( \frac{2}{e_1 e_{N-1}} - \frac{2e_{N-1}-e_1}{{e_1^3} }- \frac{2e_1-e_{N-1}}{{e_{N-1}^3}}\right) \nonumber \\
&\simeq& -2 \left(\frac{3}{4}\right)^{3/4}(N-2) e_{N-1} \xi_{01}\xi_{1,{N-1}}\xi_{N-1,0} , \nonumber \\
\end{eqnarray}
where we have neglected terms with $1/E_{N-1}$, which tends to zero when the highest excited state energy tends to infinity.

For further simplification, we use the sum rule $(m,p)=(0,0)$,
\begin{equation}\label{00SRNormalized}
e_{10}\left|\xi_{10}\right|^2 + e_{20}\left|\xi_{20}\right|^2 + \cdots + e_{N-1,0}\left|\xi_{N-1,0}\right|^2 = 1,
\end{equation}
whence
\begin{equation}\label{00SR-2}
\left|\xi_{N-1,0}\right| = \sqrt{\frac{1-\left(N-2\right)\left|\xi_{10}\right|^2}{e_{N-1,0}}}.
\end{equation}
$(m,p)=(1,1)$ gives,
\begin{equation}\label{11SR}
e_{01}\left|\xi_{10}\right|^2 + e_{21}\left|\xi_{21}\right|^2 + \cdots + e_{N-1,1}\left|\xi_{N-1,1}\right|^2 = 1
\end{equation}
whence
\begin{equation}\label{11SR-2}
\left|\xi_{N-1,1}\right| = \sqrt{\frac{1+\left|\xi_{10}\right|^2}{e_{N-1,1}}} \simeq \sqrt{\frac{1+\left|\xi_{10}\right|^2}{e_{N-1,0}}}.
\end{equation}
Introducing Eqs. (\ref{00SR-2}) and (\ref{11SR-2}) into Eq. (\ref{betamax}) we find
\begin{eqnarray}\label{betamax2}
\beta_{int}^{max} &=& -2 \left(\frac{3}{4}\right)^{3/4}(N-2) \xi_{10} \nonumber \\
&\times& \sqrt{1-(N-3)\left|\xi_{01}\right|^2 - (N-2)\left|\xi_{01}\right|^4} .
\end{eqnarray}

To optimize $\beta$ with respect to $\xi_{01}$ we first find $\xi_{01}^{max}$ by solving $\frac{\partial \beta}{\partial \xi_{01}} = 0$, leading to
\begin{equation} \label{ximax}
\xi_{01}^{max} = \left( \frac{\sqrt{\left(N-3\right)^2 + 3 \left(N-2\right)} - \left(N-3\right)}{3 \left(N-2\right)}\right)^{1/2} ,
\end{equation}
which holds for $N \geq 3$. Inserting Eq. (\ref{ximax}) into (\ref{betamax2}) results in
\begin{eqnarray}\label{betamax3}
\beta_{int, SR}^{max}&& = \left(\frac{1}{12}\right)^{1/4} \times\nonumber \\
&& \left(\left(N-2 \right)\frac{N^2 -3 - \left(N-3\right) \sqrt{N \left(N -3\right) + 3}}{(N-3) + \sqrt{N(N-3)+3}}\right)^{1/2} , \nonumber \\
\end{eqnarray}
where the subscript SR indicates the result is constrained by the sum rules.  According to Eq. (\ref{betamax3}), for $N=3$, $\beta_{int}=1$ and for $N=4$, $\beta_{int} = 1.28$ which is in agreement with the previous calculations.

The consequence of Eq. \ref{betamax3} is that $\beta_{int}$ does not converge for an infinite number of states, but diverges as
\begin{equation}\label{NinfiniteLimit}
\lim_{N \rightarrow \infty} \beta_{int,SR}^{max} =  \frac {3} {2} \left(\frac{1}{12}\right)^{1/4} \sqrt{N} .
\end{equation}
Therefore, using only the sum rules without additional constraints leads to a many-state catastrophe for the hyperpolarizability of a quantum system. The same conclusion can be made for the second hyperpolarizability, $\gamma$. This result contradicts all studies based on direct solution of the Schr\"odinger equation, as well as experiment.

\section{Discussion}

It is not surprising that previous numerical simulations did not see this behavior given the extreme/unphysical conditions that are required.  The many state catastrophe can be understood by considering the generalization from a three- to a four-level ansatz, which adds an additional degree of freedom, leading to two parameters, $E$ and $F$. The number of degrees of freedom increases in proportion to the total number of states, making it possible to find a specific combination of parameters that yields an ever-larger nonlinear response. However, since only one set of parameters out of many possible configurations gives $\beta_{int}>1$, even in millions of iterations in Monte Carlo simulations, the outliers are missed.\cite{kuzyk08.01} However, the very special set of parameters leading to Eq. (\ref{betamax3}) may not correspond to any {\em real} system, as we describe below.  Given that analytical optimization of the hyperpolarizability using the Shr\"{o}dinger equation always gives a result smaller than what is predicted by the three-level model, the three-level ansatz appears to shadow a fundamental principle that constrains the magnitude of the true nonlinear response.

\begin{figure}
  \includegraphics{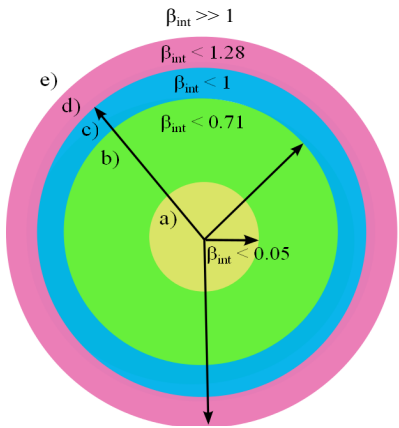}\\
  \caption{The range of attainable hyperpolarizabilities of a) real systems such as atoms and molecules; b) Hamiltonian-based approaches; c) sum rule-based Monte Carlo simulations; d) 4-level model sum-rule-constrained optimization; and e) sum rule-based $N > 4$-level model optimization.}\label{fig:circ}
\end{figure}

Fig. \ref{fig:circ} summarizes what we know about the space of all allowed transition moments and energies derivable from the sum rules.  The observations are,
\begin{itemize}
  \item All theoretical and numerical approaches that are based on exact solutions of the Schr\"odinger equation based on the standard Hamiltonian given by Eq. (\ref{Hamiltonian}) -- as well as generalizations that include spin, vector potentials, spin-orbit coupling, etc. -- lead to $\beta_{int} \leq 0.71$
  \item Monte Carlo simulations that are based on random sampling of transition moments and energies constrained to obey the sum rules, generate $\beta_{int} < 1$.
  \item The fundamental limit calculations assert that when the maximum hyperpolarizability is attained, only three energy eigenstates contribute to the nonlinear response.  This is observed for all computations using solutions of the Schr\"odinger equation when $\beta_{int} \approx 0.7$.   However, in the case of highly degenerate states, we observe that $\beta_{N\mbox{-}level}^{max} > \beta_{N-1 \mbox{-} Level}^{max} > \cdots > \beta_{3\mbox{-}Level}^{max}$ and when the number of states, $N$, approaches infinity, the sum rules impose no limit on the hyperpolarizability unless a finite $N$-level ansatz is imposed.
\end{itemize}

Motivated by the above observations, we seek to answer the following questions:
\begin{enumerate}
\item{Why does the sum rules-based Monte Carlo approach lead to different results than direct solution of the Hamiltonian?}
\item{Why have sum-rule-based Monte Carlo simulations not observed violations of the three-level anzatz, i.e. generate values with $\beta_{int} > 1$?}
\item{If the intrinsic nonlinear-optical response of an optimized N-level system can in principle be infinite, is it possible that standard Hamiltonians may be fine-tuned to lead to ultralarge hyperpolarizabilities?}
\item{Is there a well-defined limit for the largest attainable $\beta$ of a quantum systems?}
\item{How can we interpret the apparent success of the three-level ansatz and the large gap between the three-level-ansatz-based fundamental limit and most molecular systems?}
\end{enumerate}
The answer to these questions requires a detailed analysis of the theory of fundamental limits.

\subsection{Potential Issues with the Monte Carlo approach}
The intrinsic first and second hyperpolarizabilities are functions of normalized transition moments, $\xi_{ij}$'s and energies, $e_i$'s -- given by Eqs. (\ref{xi}) and (\ref{e}), respectively. In Monte Carlo simulations, these quantities are picked randomly under the constraint of the \emph{diagonal} sum rules\cite{kuzyk08.01} and then using the dipole-free formulation,\cite{kuzyk05.02,perez01.08} the hyperpolarizabilities are calculated.  The wavefunctions are assumed real and non-degenerate, so the underlying Hamiltonians that describe these system, if they exist, must be invariant under time reversal.\cite{sakur94.01}

There is potentially a pathology of this approach due to a subtle mixed-counting of states in the dipole-free expression, which uses the off-diagonal sum rules to remove the dipolar terms using the relationship,
\begin{equation}\label{eq:off-diagonal-sum-rule-dipole}
{\sum_{m=1}^\infty}^\prime \frac{\Delta x _{m0} \left| x_{m0} \right|^2} {E_{m0}} = - {\sum_{m=1}^\infty}^\prime {\sum_{n \neq m}^\infty}^\prime \frac {E_{nm} + E_{n0} } {E_{m0}^2} x_{0m}x_{mn}x_{n0} ,
\end{equation}
where $\Delta x_{m0} = x_{mm} - x_{00}$.  Since Eq. (\ref{eq:off-diagonal-sum-rule-dipole}) holds only when all infinite number of states in the $n$ summation are included, truncating the sums in the dipole free (DF) sum-over-states (SOS) expression implicitly truncates Eq. (\ref{eq:off-diagonal-sum-rule-dipole}), which leads to inaccuracies in the dipolar terms.  Indeed, it is found that the standard SOS and DF SOS expressions can disagree with each other even when the hyperpolarizabilities are calculated from a Hamiltonian.\cite{zhou06.01,zhou07.01,zhou07.02}

The issue can be described as follows.  In the standard sum-over-states (SOS) expression, the hyperpolarizability comes from two groupings of terms; ones that are functions of the dipole moment difference $\delta x_{n0} \equiv x_{nn} - x_{00}$, and other terms that are not functions of $\delta x_{n0}$.  When the standard SOS expression is evaluated using an N-state subspace, all position matrix elements $x_{nm}$ are ignored if $n>N$ OR $m>N$.  When the sum rules are used to express the dipolar terms to sums of non-dipolar terms to get the dipole-free form of the SOS expression according to Eq. (\ref{eq:off-diagonal-sum-rule-dipole}), an N-state subspace truncates the sum, making the dipolar term potentially inaccurate.  However, it is not clear if the inaccuracy is so large as to lead to hyperpolarizabilities that exceed unity.

Another potential issue originates in the Monte Carlo approach, which may allow for unphysical behavior due to the procedure of demanding agreement with truncated sum rules. For example, there may be combinations of matrix elements and energies that obey the sum rules, yet are not derivable from a Hamiltonian.  As a case in point, it is possible to contrive a system with a finite number of states that exactly obeys the sum rules;\cite{kuzyk08.01,shafe10.01} but, the nature of a system with a finite number of states renders it incompatible with being a solution of a Hamiltonian that depends on continuous functions.

However, under most circumstances, such-finite state models that obey the sum rules exactly may accurately approximate real systems with an infinite number of states if the higher-energy eigenstates do not contribute substantially to the nonlinear optical response.  In such cases, the finite-state Monte-Carlo-constrained parameters may approximate the conditions that allow finite-state models to be a good approximation to systems with an infinite number of states.\cite{kuzyk06.03} The fact that millions of Monte-Carlo runs miss the configuration that leads to the multi-state catastrophe suggests that perhaps the pathologies are rare and can be ignored.

It must be stressed that in finite state models, all the sum rules are not obeyed, since the $(N,N)$ sum rule is self contradictory and thus ignored.\cite{kuzyk06.01}  Recall that in the Monte Carlo simulation, only the diagonal sum rules up to $(N-1,N-1)$ are used to avoid the problem.  However, in the special case when all states are nearly degenerate and the highest-energy eigenstate tends to infinite energy, the system may be contrived in just the right way to violate the higher-level sum rules.  Thus, the combination of taking an unphysical configuration of states, and then taking the limiting case of an infinite number of states leads to the divergence.

It may seem paradoxical that in the limit of an infinite number of states, where the sum rules become exact, the many-state catastrophe is observed.  However, at issue is the method of how the limit is defined.

One can ask why the calculation of the fundamental limit, based on the three-level model, appears to work so well when comparing it with all hyperpolarizabilities derived from the Schr\"{o}dinger equation.  The origin of its success may lie in the fact that the three-level ansatz in effect sidesteps the infinite-state catastrophe by allowing only the minimum number of states.  The only resulting problem, based on many solutions of the Schr\"odinger equation using many different approaches,\cite{zhou07.02,kuzyk06.02,watkins09.01,szafr10.01,watkins11.01,ather12.01} is that the calculation overestimates the observations by 30\%. It is interesting that a condition that would yield the observed limit is a 2.2 level ansatz viz. Eq. (\ref{betamax3}). Since the 2-level model is unphysical, the three-level model is a compromise that yields the best result.

\subsection{Fundamental limits}
Eq. (\ref{betamax3}) represents the maximum attainable intrinsic hyperpolarizability that diverges as the number of states approaches infinity. This would suggest that for real systems with an infinite number of states, the sum rules impose no limit on the maximum hyperpolarizability. However, no real quantum system has been observed with this characteristic -- all the experimental data and numerical calculations of $\beta_{int}$ and $\gamma_{int}$ \cite{kuzyk00.02} values fall well below the limit predicted by the three-level model; and, the three-level ansatz is obeyed in the neighborhood of the limit.

The reason for this conflict may reside in the fact that sum rules permit energies and transition moments that are unphysical.  The sum rules hold for any system of particles of mass $m$, provided that the Hamiltonian obeys,
\begin{equation}\label{Hxx}
\left<p\left|\left[x,\left[H,x\right]\right]\right|q\right> = \frac{\hbar^2}{m}\delta_{pq} ,
\end{equation}
where $\left|p\right>$ and $\left|q\right>$ are eigenstates of the Hamiltonian $H$. Hamiltonians that obey Eq. (\ref{Hxx}) include those of the form,
\begin{equation}\label{sophisticatedH}
H = f(p,x) + g(x) + h(p) + k(A,B,..)
\end{equation}
where $p$ and $x$ are the momentum and position operator and $A$ and $B$ are any other operators, such as the angular momentum or spin. It is straightforward to show that a particle in the presence of electromagnetic field obeys this form, that is
\begin{equation}\label{EMHamiltonian}
H = \frac{\left(p - eA(x)/c\right)^2}{2m} + e\phi(x),
\end{equation}
where $A(x)$ and $\phi(x)$ are the vector and scalar potentials, respectively, with $f(p,x) = (p - e A(x)/c)^2$, $ g(x) = e\phi(x)$ and $h(p) = k(A,B)= 0$.

There are clearly Hamiltonians that obey Eq. (\ref{sophisticatedH}) that are more general than Eq. (\ref{EMHamiltonian}).  Perhaps one of these exotic Hamiltonians may break the $\beta_{int} < 0.71$ barrier.  Or, perhaps values larger than unity are possible.  Further study is needed to asses these possibilities.
\begin{figure}
  \includegraphics{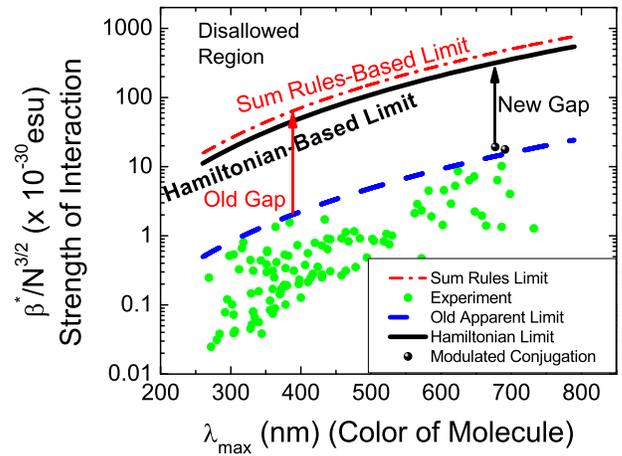}\\
  \caption{A survey of measured and calculated values of $\beta$. The red dashed-dot line is the sum rules-based limit assuming the TLA. The solid dark line is the slightly lower limit suggest by calculations using standard Hamiltonians. The green dots are experimental values reported prior to 2007. The old apparent limit, dashed blue line, was defined as the largest $\beta$ values achieved at the time. The gulf between apparent limit and the sum-rules based FL is known as the gap.}\label{fig:NewLimit}
\end{figure}

The approaches that are based on direct treatment of standard Hamiltonian are more realistic than sum-rule-based approaches that are not derivable from a Hamiltonian. The observation of the many-state catastrophe is most likely not an indictment on the sum rule calculations, but on the pathology of the highly-degenerate spectrum. In reality, we propose that the maximum attainable hyperpolarizability is most likely given by,
\begin{equation}\label{FLNew}
\beta_{max} = 0.71 \beta_{3L}^{max} .
\end{equation}

\section{A conjecture}

The sum total of all measurements and calculations using Hamiltonians leads to two observations: (1) the largest intrinsic hyperpolarizability is less than 0.709; and, (2) when a quantum system is found to have a hyperpolarizability in the vicinity of this maximum, only three states contribute.  These two observations have never been proven, so it behooves us to ask whether or not these are indeed fundamental laws of physics, or a mere coincidence.  The many-state catastrophe is an important part of the puzzle because it violates both these observations.  This is not surprising nor a matter of concern given that wave functions that are solutions to standard Hamiltonians give a range of dipole matrix elements and energies that are a subset of the dipole matrix elements and energies that obey the sum rules.

When computing limits and other fundamental relationships involving the quantum origin of optical nonlinearity, it is important to not lose sight of the fact that the hyperpolarizabilities are coefficients in an expansion of the dipole operator in terms of the applied electric field.  As such, these quantities have no meaning when a power series expansion is not possible.  The key point is that the hyperpolarizability is a quantity that is derived from perturbation theory.  Thus, any properties of a given system must be derived directly from the Hamiltonian of the quantum system including the contribution from the perturbation of the photon field.  Auxiliary quantities, such as the sum rules, must clearly be obeyed; but, as shown in this paper, may lead to behavior that is nonphysical if the properties of the Hamiltonian are not also used to constrain the system.

In the calculation of the fundamental limits of the hyperpolarizability using the sum rules, an additional auxiliary condition is added - namely the three-level ansatz.  The argument for its use is as follows. One can show rigorously and without approximation that a two-state model optimizes the polarizability, $\alpha$, in which the oscillator strength is placed into one excited state -- avoiding dilution effects that arise from spreading oscillator strength between many higher-energy states.  However, as we saw in Section \ref{sec:limits}, the sum rules show that a two-level quantum system must have $\beta = 0$ ($\beta$ is {\em minimum} when $\alpha$ is maximum).  The three-state model has the minimum number of states that leads to a physically reasonable result, and is therefore used based on the dilution argument.

While there is no rigorous proof that a three-state model yields a maximum, the approach is justified by the heuristic argument that a concentration of oscillator strength in a small number of low-energy states yields large numerators and small denominators in the SOS expression, thus maximizing the hyperpolarizability.  With more states, the oscillator strength is distributed over many states, thus diluting the nonlinearity.  However, the many-state catastrophe arises from a very specific highly-degenerate quantum system that spreads oscillator strength over many states of equal energy, thus preserving oscillator strength but without dilution because the degeneracy keeps all energy denominators small.

Before arguing that the highly-degenerate spectrum is nonphysical, one other loose end needs to be addressed -- the a priori assumption of the three-level ansatz and its potential role in being responsible for the observation that all quantum systems ever calculated obey it.  There is clearly no causal connection between calculating the limits using TLA and calculating $\beta$ for a specific Hamiltonian. The three-level ansatz is used only in the calculations of the fundamental limits, and not in the analysis of the hyperpolarizabilities that are calculated from Hamiltonians.  Thus, the observation that the three-level ansatz holds in all quantum systems tested is not tied to its assumption in the calculation of the limits.

The calculation of the fundamental limits of the polarizability is made simple by the fact that each term in the SOS expression is positive definite, and of the form,
\begin{equation}\label{AlphaTerm}
e^2 \frac { x_{0i}^2 } {E_{i0}} .
\end{equation}
As we saw in Eqs. (\ref{polarizability}) and (\ref{polarizabilityMax}), placing all the transition strength into one term, and picking the term with the minimum energy maximizes $\alpha$.  All linear harmonic oscillators are at the fundamental limit with $\alpha_{int} = 1$.  While a harmonic oscillator has many states, all the oscillator strength resides in the transition to the first excited state.  Thus, the two-state model holds exactly.  $\beta$ also is an exact two-level model for a linear harmonic oscillator, and as predicted by the sum rules, $\beta = 0$.  $\gamma$ also vanishes, as do all orders of nonlinearity by virtue of the fact that the linear harmonic oscillator is the prototypical purely linear system.

Using only the sum rules, the limits of $\beta$ are made difficult to calculate.  A typical term in the sum is of the form,
\begin{equation}\label{BetaTerm}
e^3 \frac { x_{0i} \bar{x}_{ij} x_{j0}} {E_{i0} E_{j0} } ,
\end{equation}
where $\bar{x} = x - x_{00}$.  Since each term of the form given by \ref{BetaTerm} is of indeterminant sign, the limit cannot be determined.  The many-state catastrophe shows that using only the sum rules and not truncating the SOS expression leads to a diverging result; i.e. that there is no limit.  This runs counter to the evidence.  Clearly, the correct approach is to apply the sum rules, and rather than truncating the SOS expression, to use an additional constraint that is determined from the general form of the Hamiltonian.  All attempts to find such an auxiliary condition has failed.

The state of affairs can be summarized as follows.  When applying the three-level ansatz to the SOS expression, it is found to be optimized for a specific value of energy ratio $E = E_{10} / E_{20}$  and $X = x_{10} / x_{10}^{MAX} \approx 0.76$.  The value of $\beta$ obtained in this way is found to be an upper bound for all calculations that evaluate a specific Hamiltonian when the numerical approximations that are used to evaluate the wave functions, dipole matrix, energy eigenvalues, and hyperpolarizability are small (less than $\approx$ 1\%).  In other words, all accurate calculations support these conclusions.  The best systems reach $\beta_{int} = 0.7089$.  In the vicinity of this maximum, three states dominate the second hyperpolarizability, $\beta_{xxx}$.  These suggest that the limits calculated and the three-level ansatz may be generally true.

The fact that the many-state catastrophe leads to hyperpolarizabilities that are much higher than experimentally observed and calculated values suggests that the combination of the sum rules and the three-level ansatz together yield a result that is near the true fundamental limit for hyperpolarizabilities that come from a standard Hamiltonian.

To analyze the many-state catastrophe in the vicinity of the highly-degenerate energy spectrum, the degeneracy can be lifted in a smooth way by redefining the spectrum according to
\begin{equation}\label{EpsilonSpectrum}
e_n = e_1 + \epsilon (n-1) =  1 + \epsilon (n-1),
\end{equation}
where $e_{N-1,0} = 20 \times e_{10} = 20$ for $n \neq 0,1,N-1$. Recall that all energies are normalized to $E_{10}$ so  $e_{10} = 1$.  $\epsilon$ is the splitting parameter that separates the $N-2$ degenerate states into evenly-spaced energy levels.  Eq. (\ref{EpsilonSpectrum}) can be inverted and solved for $E (= 1/e_2)$, yielding
\begin{equation}\label{EpsilonSpectrum^-1}
E = \frac {1} {1+\epsilon} .
\end{equation}

\begin{figure}
  \centering
  \includegraphics{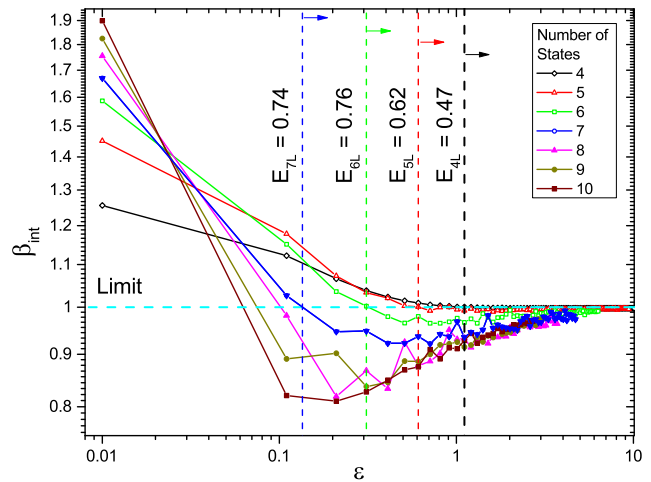}\\
  \caption{$\beta_{int}$ as a function of energy spacing parameter $\epsilon$.}\label{fig:epsilon}
\end{figure}

The nearly degenerate system is studied as follows.  For each $\epsilon$, 10,000 transition moments are randomly-sampled, and the largest value plotted in Figure \ref{fig:epsilon} for a 4- through 10-state model.  The vertical lines show where $\beta_{int} = 1$, so to the right of these lines, the hyperpolarizability is below the limit.  The subscript in the $E_{NL}$ label represents the number of states used in the calculation of $E$.  Note that this plot is approximate due to fluctuations associated with using a finite number of random samplings.

For the 4-state model, $\beta_{int} < 1$ for $\epsilon < 1.11$, which corresponds to $E=0.47$. With more states, $\beta_{int}$ falls bellow unity for smaller vales of $\epsilon$.  This is interesting in light of the fact that the best hyperpolarizabilities all share the universal value of $E \approx 0.49$, a value between that of a harmonic oscillator -- which has a polarizability at the limit, and the crossing point of $E=0.47$ for the 4-level model.

The three-level ansatz can be cast in the form
\begin{equation}\label{BetaTerm}
\beta_{int} = f(E) G(X) ,
\end{equation}
where $E = E_{10} / E_{20}$ and $X = x_{10} / x_{10}^{MAX}$.  Note that this expression is postulated to hold only near the fundamental limit and is observed to hold in the vicinity of a local maximum of $\beta_{int}$ when close to the limit.  The fundamental limit is found for $f(0) = 1$ and $G(\sqrt[-4]{3} \approx 0.76)$.  In potential optimization studies, the class of optimized potentials all share the universal properties of $\beta_{int} = 0.7089$, $E =0.49$, and $X = 0.79$;\cite{zhou07.02} so, $X$ is near the value needed to be at the limits.  As such, the hyperpolarizability appears to be limited by the nature of the energy spectrum, as suggested in studies of various organic molecules.\cite{tripa04.01, tripa06.01}

In the calculations of the fundamental limit, $X$ and $E$ are assumed to be independent and are thus separately optimized.  Since the universal value of $X$ is near optimum, we focus on $E$.  For a harmonic oscillator, $E = 1/2$ and $f(1/2) = 1/\sqrt{2} = 0.707$, a value tantalizing close to the universal value of $\beta_{int}$.  This suggests that perhaps $X$ and $E$ are not independent, so the best values of the intrinsic hyperpolarizability are constrained to have an energy spectrum that is similar to the harmonic oscillator when the transition is nearly optimized.

The sum rules (which are more general than what one obtains from solving the standard Schr\"odinger equation), the three-level ansatz (which is not generally true based on the exception found using the many-state catastrophe), and the assumption that $E$ and $X$ are independent (unproven) yields a calculated fundamental limit that is within 30\% of the maximum value observed in many classes of optimized potentials.  Furthermore, universal values are found that add credence to the ideas of absolute limits of scaled hyperpolarizabilities; and, the best measured molecules, as shown in Figure \ref{fig:NewLimit}, scale parallel to the limit lines -- another indication that there is some substance behind the results.

We propose that the observations that the state vectors associated with the many-state catastrophe live in the realm obeyed by the sum rules but beyond solutions of the Schr\"odinger equation.  Thus, in calculating limits of real systems, we must exclude those cases.  Figure \ref{fig:epsilon} illustrates the demarcation between the allowed (below $\beta_{int} = 1$) and disallowed. A four-level model demands that the degeneracy parameter be $\epsilon \geq 1.11$ yielding $E \leq 0.47$; a five-level model demands that $\epsilon \geq 0.61$ yielding $E \leq 0.62$, etc., for a system of equally spaced intermediate states.  Clearly, there are many other possible types of energy spectra, but this example illustrates how one can define a condition that sets limits on the allowed spectra for quantum systems that obey the standard Schr\"odinger equation.

Given these results, we propose the following conjectures:
\begin{center}\line(1,0){150}\end{center}
\begin{enumerate}[I.]
\item The three-level ansatz is true.
\item $\beta_{int} \leq 0.7089$ for any system derivable from a standard Hamiltonian.
\end{enumerate}
\begin{center}\line(1,0){150}\end{center}

The SOS expression for $\beta$ has terms of the form Eq. (\ref{BetaTerm}), which are indeterminate in sign, making the calculations of the limit impossible without the use of of an auxiliary condition.  Without placing a constraint on the type of potentials that are allowable, the limit is not calculable.  As such, it is possible that these conjectures are true but unprovable.

As a corollary to our conjectures, we propose that they are indeed true but unprovable.  The fact that the conjecture has not been proven is not evidence that it is unprovable.  However, without an additional condition, the conjectures can not be proven.  As such, we must wait until such a condition is found.

\section{Conclusion}

The many-state catastrophe is an observation that calls into question the assumptions used in calculating the fundamental limit.  When many states are arranged into a highly-degenerate energy spectrum, the hyperpolarizability diverges in the limit of an infinite number of states.  While the divergence can be avoided by arguing that real systems that obey a standard Hamiltonian cannot have such a spectrum, the fact that this outlier both obeys the sum rules and contradicts the assumptions in the calculation of the limits demands that the assumptions be revisited.

By smoothly splitting the degeneracy using an energy spacing parameter, the fundamental limit theory is found to hold when $E>0.47$, where the many state catastrophe is avoided.  The fact that the observed universal value for Hamiltonians expressed in terms of a potential energy function near the largest observed upper bound of the hyperpolarizability ($E \approx 0.49$) meets this condition supports the assertion that systems representable by a potential energy function have a restricted energy spectrum of this form.

Based on these observations, we propose a conjecture that the three-level ansatz is a fundamental law and that the true fundamental limit is given by $\beta_{int} = 0.7089$.  We also posit that while true, the conjectures may not be provable due to the difficulty (or impossibility) of defining an auxiliary condition that standard Hamiltonians must obey while excluding more general systems.

The three-level ansatz appears to be the correct auxiliary condition needed to calculate the the true fundamental limit to within 30\% of the correct value.  With minor modifications, the true limit can be exactly calculated. Reconciling the three-level ansatz with the many-state catastrophe and the overestimation of the true limit are open research problems under study.

\begin{acknowledgments}
We would like to thank the National Science Foundation (ECCS-1128076) for generously supporting this work, and Sean Mossman for his useful suggestions.
\end{acknowledgments}

\end{document}